\documentclass[final,authoryear,5p,times,twocolumn]{elsarticle}%
\pdfoutput=1%
\usepackage[utf8]{inputenc}%
\usepackage[english]{babel}%
\usepackage{graphicx}%

\journal{Quaternary International}

\begin{document}

\begin{frontmatter}

\title{Reconstructions in human history by mapping dental markers in living
Eurasian populations}

\author[yunc]{V.\,F.\,Kashibadze\corref{cor1}}
\ead{verdari@gmail.com}

\author[sao]{O.\,G.\,Nasonova}

\author[sao]{D.\,S.\,Nasonov}

\cortext[cor1]{Corresponding author. Southern Scientific Centre, Russian Academy
of Sciences, 41 Chekhov Street, Rostov-on-Don, 344006, Russia}

\address[yunc]{Southern Scientific Centre, Russian Academy of Sciences, 41
Chekhov Street, Rostov-on-Don, 344006, Russia}

\address[sao]{Special Astrophysical Observatory, Russian Academy of Sciences,
Nizhniy Arkhyz, Karachai-Cirkassian Republic, 369167, Russia}

\begin{abstract}
On the base of advantages in gene geography and anthropophenetics the
phenogeographical method for anthropological research is initiated and
experienced using dental data. Statistical and cartographical analyses are
provided for 498 living Eurasian populations. Mapping principal components
supplied evidence for the phene pool structure in Eurasian populations and for
reconstructions of our species history on the continent. The longitudinal
variability seems to be the most important regularity revealed by principal
components analysis (PCA) and mapping proving the division of the whole area
into western and eastern main provinces. So, the most ancient scenario in the
history of Eurasian populations was developing from two perspective different
groups: western group related to ancient populations of West Asia and the
eastern one rooted by ancestry in South and/or East Asia. In spite of the
enormous territory and the revealed divergence the populations of the continent
have undergone wide scale and intensive time-space interaction. Many details in
the revealed landscapes could be backgrounded to different historical events.
The most amazing results are obtained for proving migrations and assimilation as
two essential phenomena in Eurasian history: the wide spread of the western
combination through the whole continent till the Pacific coastline and the
envision of the movement of the paradox combinations of eastern and western
markers from South or Central Asia to the east and to the west. Taking into
account that no additional eastern combinations in the total variation in Asian
groups have been found but mixed or western markers’ sets and that eastern
dental characteristics are traced in Asia since Homo erectus, the assumption is
made in favour of the hetero-level assimilation in the Eastern province and of
net-like evolution of our species.
\end{abstract}


\end{frontmatter}

\section{Introduction}
\label{intro}

This study shoots out from two inspiring sources: progress in genetic
reconstructions of our species history and experience in accumulating and
analysis rich dental data on living and fossil human populations. The innovation
is in combining both advantages to provide new knowledge on Eurasian ancestry.
Principles of anthropophenetics determine the basic approach to the research and
allow using methods of population genetics. Phenes (discrete irreducible
morphological traits) yield to genes in number and in marking precision but win
in extent of genome covering. Dental traits provide the best possibility to
examine directly time records in populations. Mapping applications can detect
many different patterns hidden in numerous tabled data; each pattern seems to
have a certain historical content. Computer maps provide both analysis and
visualization of the enormous volume of data accumulated in dental anthropology.
The study is the first experience of this sort.

\section{Material and methods}
\label{data}

The study involves data from 498 samples, 50257 individuals in total, drawn from
living populations in Eurasia and Africa.

The material was taken from a great number of publications, the major its part
is presented in two generalizing books by \citeauthor{Z73} (\citeyear{Z73},
\citeyear{ZK89}); the other part is our own recent data on populations of the
Caucasus (86 samples), the Far East (3 samples) and on the Russians (27 samples)
\citep{K06}.

To process this rich information the universal system of analysis, visualization
and mapping of dental data {\slshape{}Eurasia} has been developed by the
authors. All dental data on living and fossil Eurasian populations available to
this moment are managed by the MySQL relational database. Data refer to 830
populations, 32 dental traits, no less than 120 dental phenes as several grades
or discrete variations of a trait, and 12 historical periods from the
Palaeolithic to the present.

All the data operations including statistics and visualization are implemented
as routines written in Python. In several cases C modules are used to improve
the performance. The basic statistics are executed using the principal component
analysis (PCA) method with the help of PCA Module algorithms
(http://folk.uio.no/henninri/pca\_module/). Mapping of separate dental markers
frequencies and PC scores is accomplished via the Matplotlib Basemap Toolkit
(http://matplotlib.sourceforge.net/basemap/doc/html/).

The whole system is managed through web interface built within Django framework
(http://www.djangoproject.com/) allowing to handle the database, to generate
dynamic graphics and to save it in vector or bitmap formats.

The study program is common in the Russian Federation and includes 32 non-metric
dental traits (\citealp{Z68}, \citeyear{Z73}; \citealp{ZK89}). But the real
situation is that only few markers are usually presented in published tables of
frequencies, so we had to find reasonable balance between number of populations
and number of markers involved into PCA. Thus the numbers are 498 and 8
respectively.

In total, 143 phenogeographical maps have been created, but in the present short
paper only 4 of them, i.e. mapping scores of the four PCs, are overviewed and
discussed.

The maps were constructed by interpolating the PC score distribution with the
Gaussian as a weight function. We have adopted the following parameters for
constructing maps in the case of living populations: the averaging window
$\sigma = 1.0^\circ$, the weight function range $6.1^\circ$ and the total number
of grid knots 50,400. Small black points indicate the location of the
populations under investigation. Only aboriginal groups are investigated.

\section{Results and discussion}
\label{results}

Conclusions derived from maps interpretation are often ultimately compiled on
the basis of how authors envisage their data fit with established genetic,
archaeological or linguistic theories. We try to make such conclusion in the
most independent way and on the basis of experience in dental anthropology and
anthropophenetics. Only this approach can provide really new knowledge.

The PCA results are presented in table~\ref{table-pca}.

\begin{table}
\caption{The loadings of 8 dental traits in the first 4 PC scores.}
\label{table-pca}
\begin{center}
{\footnotesize{}\tabcolsep=0.8ex%
\begin{tabular}{lcccc}
\hline
 Dental phenes & 1$^\mathrm{st}$ PC & 2$^\mathrm{nd}$ PC & 
                 3$^\mathrm{rd}$ PC & 4$^\mathrm{th}$ PC \\
\hline
 {\itshape{}Variability} & 53\,\% & 13\,\% & 9\,\% & 7\,\% \\
\hline
Shoveling $I^1$:                          & $-$0.40 & $+$0.01 & $-$0.23 & $+$0.11 \\
4-cusped $M_1$                            & $+$0.25 & $+$0.62 & $-$0.55 & $+$0.28 \\
6-cusped $M_1$                            & $-$0.39 & $-$0.19 & $-$0.02 & $+$0.03 \\
4-cusped $M_2$                            & $+$0.41 & $+$0.24 & $+$0.00 & $-$0.05 \\
Distal trigonid crest on $M_1 $           & $-$0.33 & $+$0.37 & $-$0.13 & $-$0.76 \\
Deflecting wrinkle on $M_1$               & $-$0.39 & $-$0.08 & $-$0.32 & $+$0.48 \\
Confluence furrows 2\,med and II on $M_1$ & $+$0.36 & $-$0.28 & $+$0.13 & $+$0.03 \\
Carabelli cusp                            & $+$0.24 & $-$0.55 & $-$0.72 & $-$0.31 \\
\hline
\end{tabular}}
\end{center}
\end{table}

The weight of a certain trait is defined as loadings for a corresponding
normalized $\varphi$-transformed (arcsine-transformed) frequency in the linear
combination specifying the component. PCA was applied to the among-group dental
variability.

\begin{figure*}[hp]
\centering
\includegraphics[width=18.35cm]{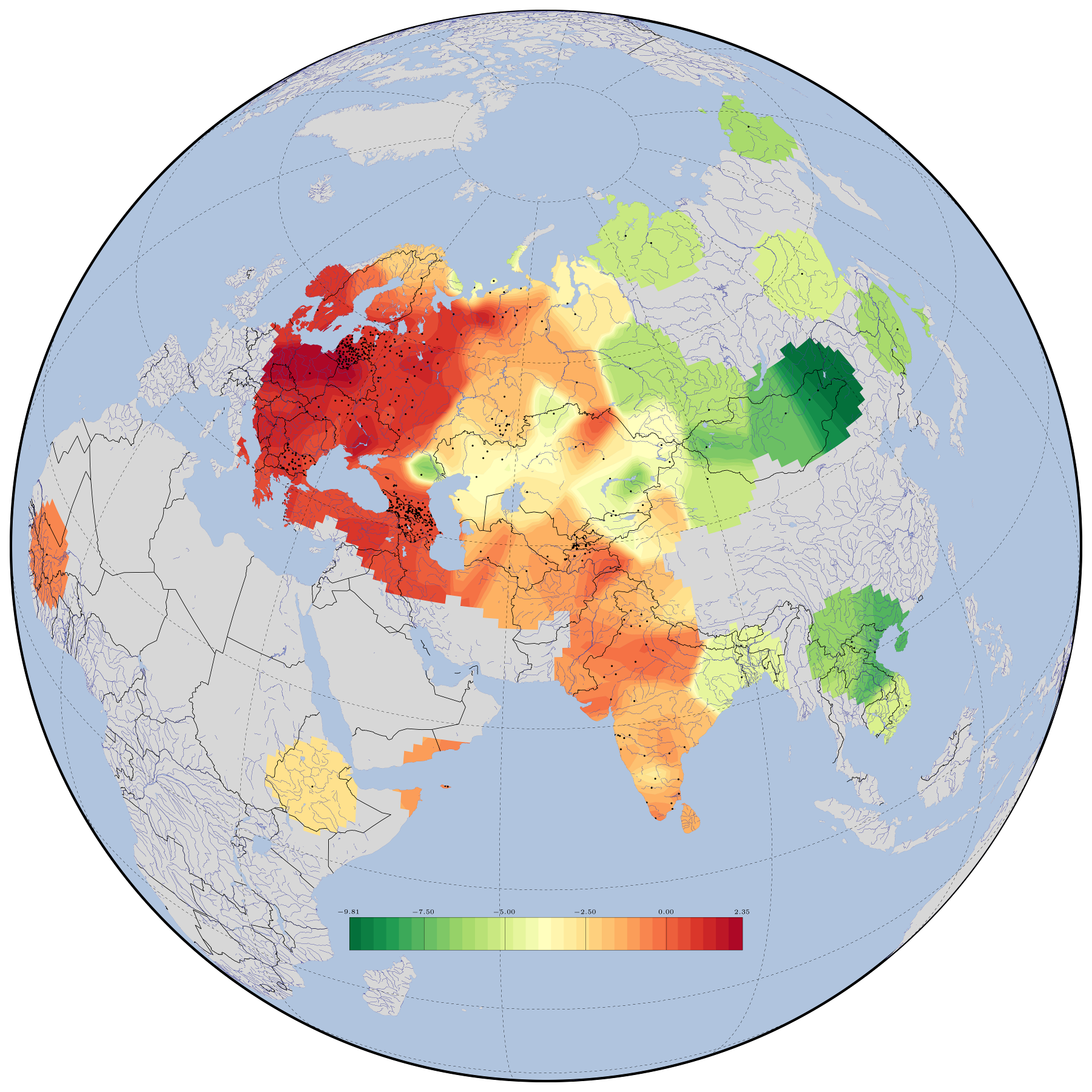}
\caption{ The distribution of the 1$^\mathrm{st}$ PC score in living Eurasian
populations.}
\label{1pc}
\end{figure*}

The longitudinal variability of phene pool in Eurasian populations seems to be
the most important regularity revealed by mapping and PCA (fig.\,\ref{1pc}). The
geographical factor provides the main contribution to the revealed diversity.
The 1$^\mathrm{st}$ PC explains 53\,\% of the total phenetic variation. All
populations under investigation are divided into two main provinces: the western
area with high PC1 scores and the eastern one with low scores.

Several scenarios of different time series could determine this pattern. The
larger is the space embracing populations that share similar frequencies, the
deeper is the time in their divergence. So, we can suggest the most ancient
scenario in the history of Eurasian populations was developing from two
perspective different groups.

Africa presented by populations from Ethiopia \citep{SNKZ84} and the Republic of
Mali \citep{K77} joins the western province.

The map shows clines of evident phene flows from Near East north-eastward to
Siberia. It could be backgrounded to intensive post-Neolithic expansion or to
any earlier events. It’s the matter for further research. Another flow can be
easily traced from east to west along the steppe belt of the continent. It is
explained by the latest (early medieval) expansion from Inner Asia evoking
oscillatory migratory waves in population settled along the steppe belt, thus
comprising a complicated system in populational interaction. The Kalmyks in the
south of East Europe is the western final point in this expansion.

The contact zone between the provinces occupies the Urals, West Siberia, Middle
Asia and India.

Regarding the split in two main provinces it should be noted that this
phenomenon in Eurasia can be traced since Homo erectus. Indeed, archaic western
forms show a low grade of shoveling and poor differentiation in odontoglyphical
patterns on molars versus extremely developed shoveling and richness in
odontoglyphics in the eastern province (\citealp{ZK89}, pp.\,196--197).
Chronology and dynamics in different morphological systems’ evolution seem to be
rather independent, while dental characteristics demonstrate much antiquity in
phylogeny and provide a direct bridge from the present to the far past
inaccessible for cranial traits, as our recent research on ancient and living
Caucasian populations showed \citep{K06}. It is worth to mention that both
western and eastern forms of Homo erectus had five-cusped and six-cusped lower
molars \citep{ZK89}, their gracilization is a peculiar characteristic of Homo
sapiens, still eastern living populations keep higher frequencies in
five-six-cusped lower molars. It is difficult to ignore these most important
data provoking an assumption of the replacement in hominines in the west of the
continent and of the hetero-level assimilation in the Eastern province.

\begin{figure*}[hp]
\centering
\includegraphics[width=18.35cm]{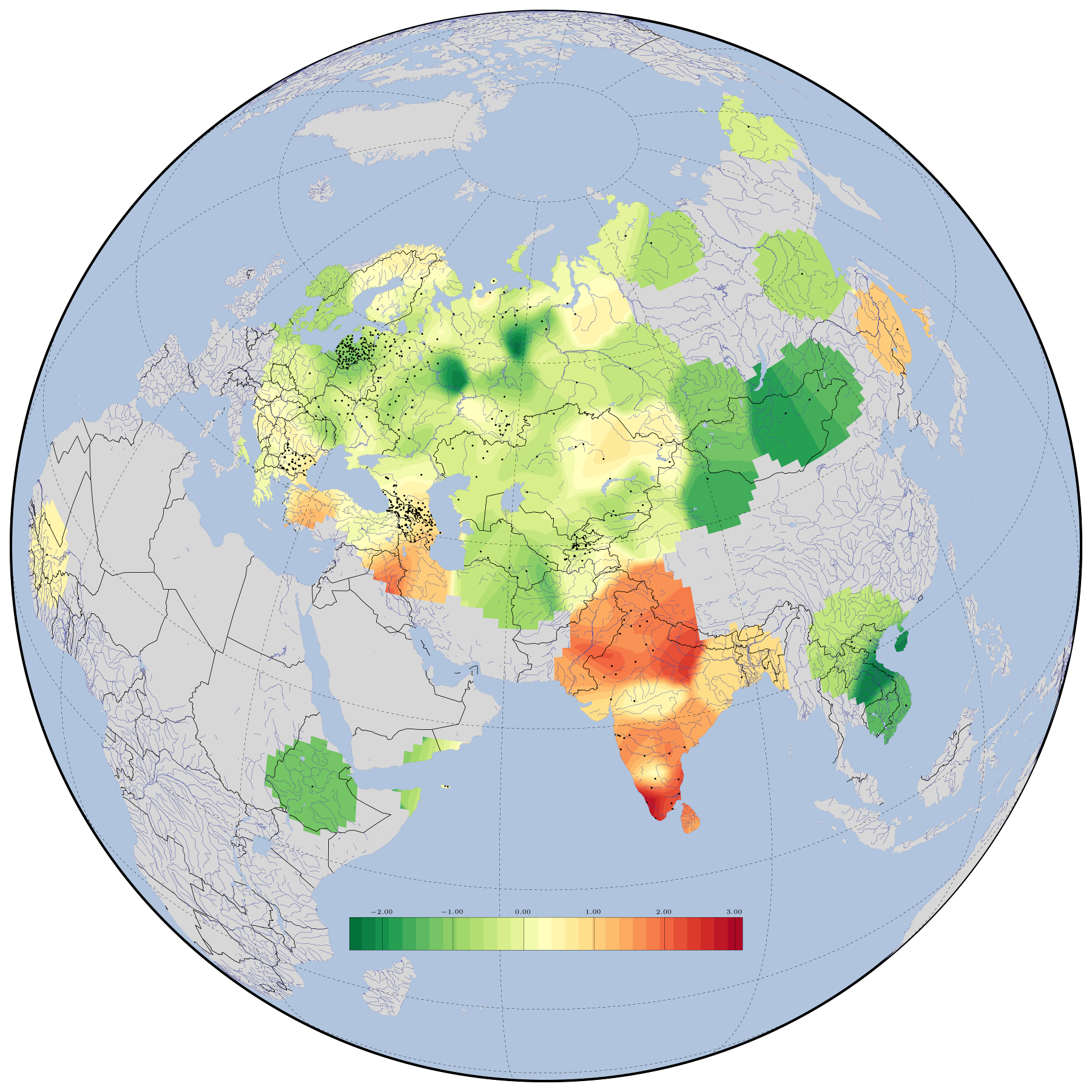}
\caption{ The distribution of the 2$^\mathrm{nd}$ PC score in living Eurasian
populations.}
\label{2pc}
\end{figure*}

The 2$^\mathrm{nd}$ PC explains 13\,\% of the total phenetic variation
(fig.\,\ref{2pc}). The map shows more the latitudinal variability of phene pool
in Eurasian populations. The scores of the PC are high in Dravidian and Munda
groups of India, in other Indian and some Far East populations as well as in
many populations in the south of West Asia and in the north of Europe and
Siberia. In fact, the 2$^\mathrm{nd}$ PC presents the paradox combination of
eastern (the distal trigonid crest on the first lower molar) and western markers
(four-cusped lower molars, precisely the first one). In our previous study on
the Caucasian populations we suggested both southern and northern gracile
subsets in West Eurasia had developed from one ancestral eastern group
\citep{K06}. The pattern on the map supports this assumption. For the first time
we find the traces of the initial group in the east province. We can envisage
the movement of this ancient group from South Asia to the east and to the west,
subsequent splitting the west flow into northern and southern subsets, probably
as a result of populating postglacial continental space.

\begin{figure*}[hp]
\centering
\includegraphics[width=18.35cm]{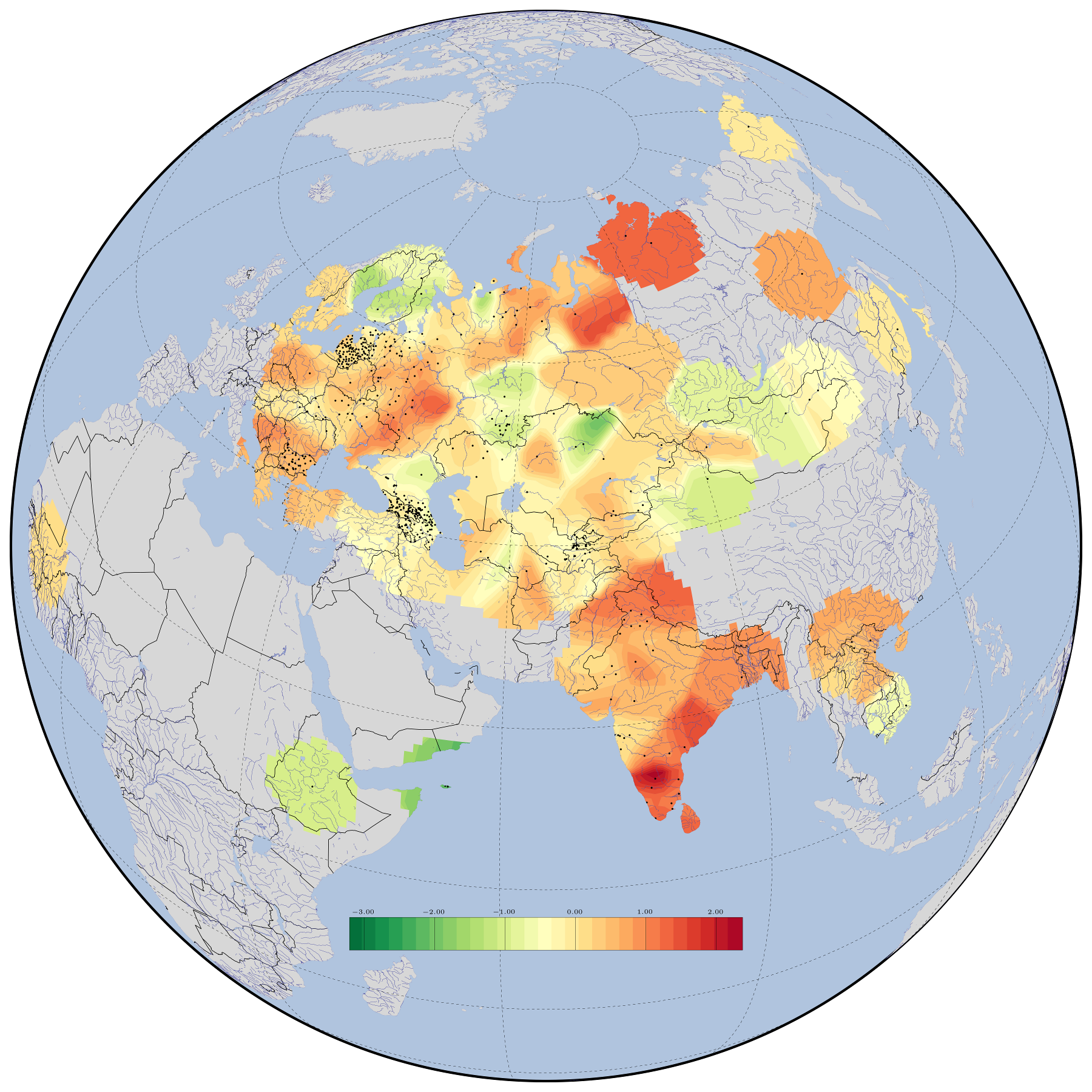}
\caption{ The distribution of the 3$^\mathrm{rd}$ PC score in living Eurasian
populations.}
\label{3pc}
\end{figure*}

The 3$^\mathrm{rd}$ PC explains 9\,\% of the total phenetic variation
(fig.\,\ref{3pc}). High scores of the 3$^\mathrm{rd}$ PC are determined by
loadings mainly in western traits (Carabelli cusp and four-cusped lower first
molars). We've revealed this combination for west living and fossil populations
of the continent in our previous investigation of the Caucasus in the
anthropohistorical space of Eurasia \citep{K06}.

The wide spread of this component through the whole continent seen in the
map~\ref{3pc} was an absolutely unexpected and amazing phenomenon. It provides
evidence for wide scale ancient populating the continent by a group or close
groups mainly from west to east. Maybe this event could happen at different
times and maybe repeatedly. 

\begin{figure*}[hp]
\centering
\includegraphics[width=18.35cm]{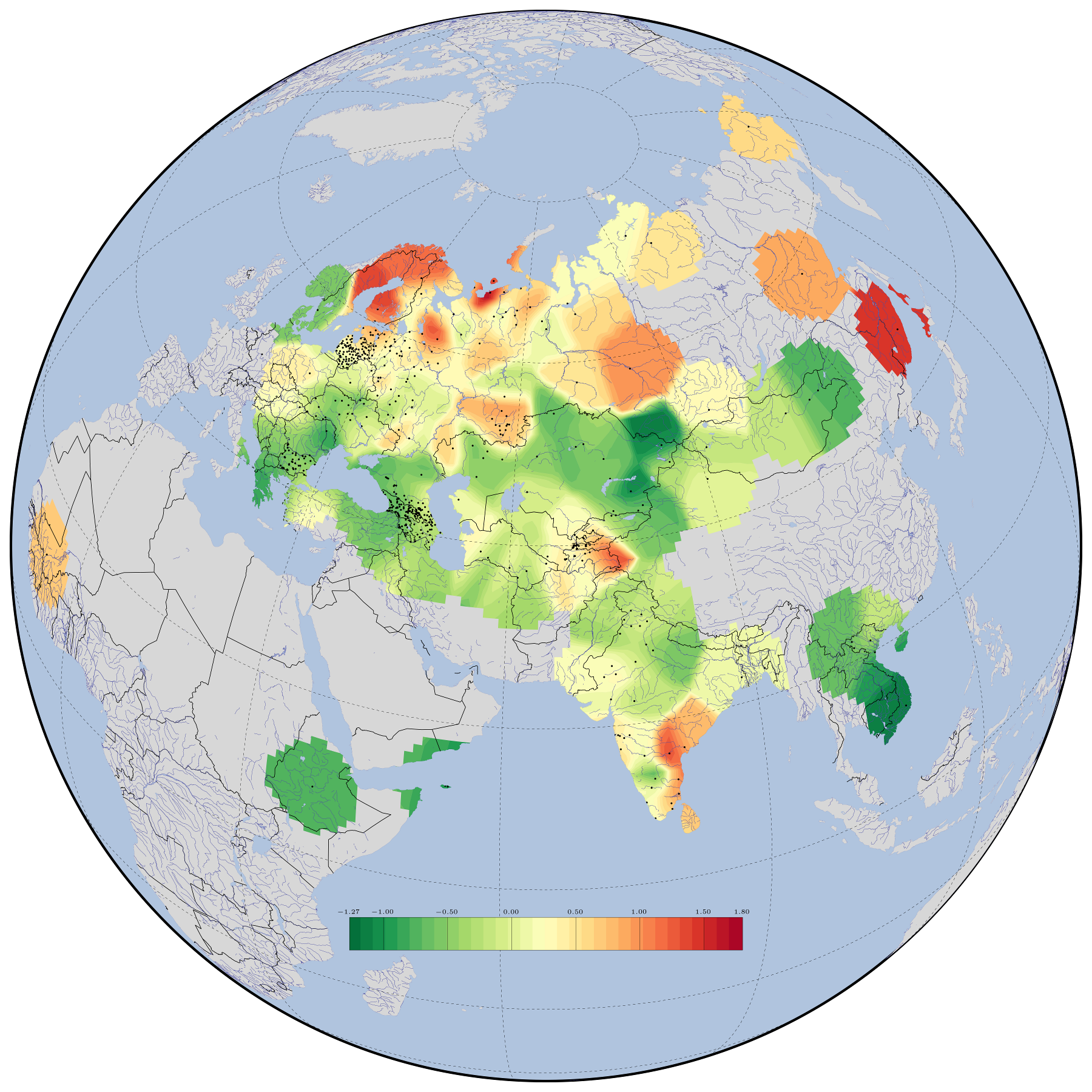}
\caption{ The distribution of the 4$^\mathrm{th}$ PC score in living Eurasian
populations.}
\label{4pc}
\end{figure*}

The 4$^\mathrm{th}$ PC describes 7\,\% of the total phenetic variation. This PC
presents again a combination of eastern and western traits (distal trigonid
crest and Carabelli cusp), and it covers south regions of Eurasia. The revealed
landscape more obviously divides the continent into southern and northern
halves. The northern zone is occupied by another mixed combination (deflecting
wrinkle and four-cusped first lower molars) common to Finn-Ugric populations
(\citealp{Z73}; \citealp{ZK89}). The map (fig.\,\ref{4pc}) again detects hidden
patterns and more wide traces of this combination in southern, central and
eastern areas of the continent. The highest scores of the last combination are
found in some marginal coastline and in central mountain populations of Eurasia.
The map traces the dispersal of a branch of this combination to the west via the
Middle Urals and its subsequent irradiation in the European territory.

It must be emphasized that two of the four PCs are composed by combinations of
western and eastern markers. Diminishing eastern traits frequencies at different
grades in different groups of west continental populations \citep{K06} and
worldwide in shoveling \citep{M85} is a discovered phenomenon. So, we can
suggest that ancestral polymorphic or assimilated populations should have even
more expressed eastern component. Evolutionary factors, including genetic drift,
selection and gene flows, may have altered the patterns of phenetic frequency
and distribution in existing populations.

However, the time depths of the revealed landscapes are still not known exactly,
so associating them with particular historic and demographic events seems to be
speculative at the moment. To provide clear dating additional studies in
integrating with established genetic, archaeological and linguistic evidence
should be launched. Since dental markers provide the best possibility to examine
directly time records in populations, an alternative perspective is in phenetic
investigations of fossil Eurasian groups. East Europe and adjacent areas rich in
fossil data seem to be the region to start with.

\section{Conclusions}
\label{concl}

In spite of the enormous territory and the revealed divergence the populations
of the continent have undergone wide scale and intensive time-space interaction.
The maximal phenetic diversity was detected in India, respectively lesser in
North Europe, West Siberia and Near East. Many details in the revealed landscape
could be backgrounded to different historical events.

The maps visualize the most important results in analysis: the wide spread of
the western combination through the whole continent till the Pacific coastline
and the envision of the dispersal of the paradox combinations of eastern and
western markers from South or Central Asia to the east and to the west. Taking
into account that no additional eastern combinations in the total variation in
Asian groups have been found but mixed or western markers' sets and that eastern
dental characteristics are traced in Asia since Homo erectus, the choice between
the ancestral polymorphism and the hetero-level assimilation in the Eastern
province is made in favour of the latter.

\section*{Acknowledgements}
\label{ackn}
The study was supported by the Russian Foundation of Basic Research, grant
08-06-00124.

The authors are grateful to Prof. Ele\-na Ba\-la\-nov\-ska, Research Centre for
Medical Genetics, Russian Academy of Medical Sciences, for the initial support
of the idea of this research; and to Petr Voi\-tsik, Astro Space Center, the
Le\-be\-dev Physical Institute of the Russian Academy of Sciences, for his
valuable aid in software development; and, last but not least, to the organizers
and the participants of the 2010 meeting INQUA-SEQS for the inspiring
communication.

\bibliographystyle{elsarticle-harv}

\end{document}